\documentstyle[12pt]{article}
\newcommand\be{\begin{equation}}
\newcommand\ee{\end{equation}}
\newcommand\bea{\begin{eqnarray}}
\newcommand\eea{\end{eqnarray}}
\newcommand\ket[1]{|#1\rangle}
\newcommand\bra[1]{\langle #1|}

\newcommand{\fatalpha}{{\bf \alpha \kern -0.44em \alpha}}
\newcommand{\fatsigma}{{\bf \sigma \kern -0.54em \sigma}}
\newcommand{\tpchi}{{\bf \chi \kern -0.35em \chi}}
\newcommand{\llambda}{{\bf \lambda \kern -0.45em \lambda}}

\bibliography{plain}
\title{\bf Finite quantum tomography via semidefinite programming} \vspace{20mm}
\author{ M. A. Jafarizadeh$^{a,b,c}$
 \thanks{E-mail:jafarizadeh@tabrizu.ac.ir},
M.Mirzaee$^{a,b}$
\thanks{E-mail:mirzaee@tabrizu.ac.ir},M.Rezaee$^{a,b}$
\thanks{E-mail:karamaty@tabrizu.ac.ir}
\\
\\
$^a${\small Department of Theoretical Physics and Astrophysics,
Tabriz University, Tabriz 51664, Iran.} \\ $^b${\small Institute
for Studies in Theoretical Physics and Mathematics, Tehran
19395-1795, Iran.} \\ $^c${\small Research Institute for
Fundamental Sciences, Tabriz 51664, Iran.}} \pagebreak

\vspace{20mm}
\begin{document}
\maketitle \vspace{15mm}
\newpage
\begin{abstract}
Using the the convex  semidefinite programming method and
superoperator formalism we obtain the finite quantum tomography of
some mixed quantum states such as: qudit tomography, N-qubit
tomography, phase tomography and coherent spin state tomography,
where that obtained results are in agreement with those of
References \cite{schack,Pegg,Barnett,Buzek,Weigert}.
 {\bf Keywords: finite quantum tomograpy, Semi-definite programming,superoperator formalism, qubit quantum tomography and truncation.}
{\bf PACs Index: 03.65.Ud }
\end{abstract}
\vspace{70mm}
\newpage
\section{Introduction}
The quantum complementarity principle does not allow to recover
the quantum state from measurements on a single system, unless we
have some prior information on it. On the other hand, the no
cloning theorem ensures that it is not possible to make exact
copies of a quantum system, without having prior knowledge of its
state. Hence, the only possibility for devising a state
reconstruction procedure is to provide a measuring strategy that
employs numerous identical (although unknown) copies of the
system, so that different measurements may be performed on each
of the copies.

The problem of state estimation resorts essentially to estimating
arbitrary operators of a quantum system by using the result of
measurements of a set of observables. If this set of observables
is sufficient to give full knowledge of the system state, then we
define it a quorum. Notice that, in general, a system may allow
various, different quorums. Quantum tomography was born \cite{1,4}
as a state reconstruction technique in the optical domain, and
has recently been extended \cite{1p} to a vast class of systems.
By extension, we now denote as "Quantum Tomography" all unbiased
quantum state reconstruction procedures, i.e. those procedures
which are affected only by statistical errors that can be made
arbitrarily small by increasing the number of measurements.
Tomography makes use of the results of the quorum measurements in
order to reconstruct the expectation value of arbitrary operators
(even not observables) acting on the system Hilbert space.

In principle, a precise knowledge of the density matrix would
require an infinite number of measurements on identical
preparations of radiation. However, in real experiments one has
only a finite number of data at ones disposal, and thus a
statistical analysis and errors estimation are needed.

Authors of  Ref. \cite{Buzek} presented several schemes for a
reconstruction of states of quantum systems from measured data:

(1) The maximum entropy (MaxEnt) principle leads to a complete
reconstruction of quantum states, i.e. quantum states are
uniquely determined.

(2) Quantum systems can be estimated with the help of quantum
Bayesian inference.

(3) Estimation of a quantum state with the highest fidelity and
Showed how this optimal measurement can in principle be realized
\cite{Buzek}.

 On
the other hand, over the past years, semidefinite programming
(SDP) has been recognized as valuable numerical tools for control
system analysis and design. In (SDP) one minimizes a linear
function subject to the constraint that an affine combination of
symmetric matrices is positive semidefinite. SDP, has been studied
(under various names) as far back as the 1940s. Subsequent
research in semidefinite programming during the 1990s was driven
by applications in combinatorial optimization\cite{Luo04},
communications and signal processing \cite{Luo03,Luo02,Luo01},
and other areas of engineering\cite{Luo05}. Although semidefinite
programming is designed to be applied in numerical methods it can
be used for analytic computations, too. Some authors try to use
the SDP to construct an explicit entanglement witness
\cite{Doherty,Parrilo}. Kitaev used semidefinite programming
duality to prove the impossibility of quantum coin flipping
\cite{Kitaev}, and Rains gave bounds on distillable entanglement
using semidefinite programming \cite{Rains}. In the context of
quantum computation, Barnum, Saks and Szegedy reformulated
quantum query complexity in terms of a semidefinite program
\cite{Barnum}. The problem of finding the optimal measurement to
distinguish between a set of quantum states was first formulated
as a semidefinite program in 1972 by Holevo, who gave optimality
conditions equivalent to the complementary slackness conditions
\cite{Helestrum}. Recently, Eldar, Megretski and Verghese showed
that the optimal measurements can be found efficiently by solving
the dual followed by the use of linear programming \cite{Eldar}.
Also in \cite{Lawrence}  used semidefinite programming to show
that the standard algorithm implements the optimal set of
measurements. All of the above mentioned applications indicate
that the method of SDP is very useful.

In a laboratory and in practice, we always deal with finite
ensembles of copies of the measured system. This implies the need
of developing novel tools specially designed to process realistic
and finite experimental samples. Then it is necessary to truncate
the Hilbert space to a finite dimensional basis \cite{Vbuzek}. In
this paper we use the   SDP method in order to obtain quantum
tomography with truncating the infinite Banach space to a finite
dimensional basis.

 The paper is organized as follows:\\ In
section-2 we define semidefinite programming. In section -3 we
define superoperator formalism.  In section -4 we describe the
projection method and using SDP method and superoperator
formalism  we obtain finite quantum tomography. In section -5 we
obtain some typical finite quantum tomographic examples, such as:
finite dimensional qudit quantum tomography, N-qubit tomography,
finite dimensional phase tomography  and coherent spin state
tomography with SDP method and superoperator formalism. The paper
is ended with a brief conclusion.
\section{Semi-definite programming}\label{semi}
A SDP  is a particular type of convex optimization problem
\cite{optimize}. A SDP problem requires minimizing a linear
function subject to a linear matrix inequality (LMI) constraint
\cite{optimize1}: \be \label{sdp1}\begin{array}{cc}
\mbox{minimize} & {\cal P}=c^{T}x
\\ \mbox{subject to} & F(x)\geq 0,
\end{array}\ee where c is a given vector,
$x^{T}=(x_{1},...,x_{n}), $ and $F(x)=F_{0}+\sum_{i}x_{i}F_{i},$
for some fixed hermitian matrices $F_{i}$. The inequality sign in
$F(x)\geq 0$ means that $F(x)$ is positive semidefinite.

This problem is called the primal problem. Vectors x whose
components are the variables of the problem and satisfy the
constraint $F(x) \geq 0$ are called primal feasible points, and
if they satisfy $F(x) > 0$ they are called strictly feasible
points. The minimal objective value $c^{T} x$ is by convention
denoted as ${\cal P}^{\ast}$ and is called the primal optimal
value.

Due to the convexity of  set of feasible points, SDP  has a nice
duality structure, with, the associated dual program being: \be
\label{sdp2}\begin{array}{cc} \mbox{maximize} & -Tr[F_{0}Z] \\
    & Z\geq 0 \\  & Tr[F_{i}Z]=c_{i}. \end{array}\ee

Here the variable is the real symmetric (or Hermitean) matrix Z,
and the data c, $F_{i}$ are the same as in the primal problem.
Correspondingly, matrices Z satisfying the constraints are called
dual feasible (or strictly dual feasible if $Z > 0$). The maximal
objective value $-Tr F_{0}Z$, the dual optimal value, is denoted
as $d^{\ast}$.

The objective value of a primal(dual) feasible point is an upper
(lower) bound on ${\cal P}^{\ast}$($d^{\ast}$.  The main reason
why one is interested in the dual problem is that one can prove
that $d^{\ast} \leq {\cal P}^{\ast}$, and under relatively mild
assumptions, we can have ${\cal P}^{\ast} = d^{\ast}$. if the
equality holds, one can prove the following optimality condition
on x:

A primal feasible $x$ and a dual feasible $Z$ are optimal which is
denoted by $\hat{x}$ and $\hat{Z}$ if and only if  \be
\label{slacknes} F(\hat{x}) \hat{Z}=\hat{Z} F(\hat{x})=0. \ee This
latter condition is called the complementary slackness condition.

In one way or another, numerical methods for solving SDP problems
always exploit the inequality $d \leq d^{\ast} \leq {\cal
P}^{\ast} \leq {\cal P}$, where d and ${\cal P}$ are the objective
values for any dual feasible point and primal feasible point,
respectively. The difference \be\label{sdp3}{\cal P}-d= c^{T}x+
Tr[F_{0}Z]= Tr[F(x)Z]\geq 0 \ee is called the duality gap. If the
equality $d^{\ast}={\cal P}^{\ast}$ holds, i.e., the optimal
duality gap is zero, then we say that strong duality holds.
\section{Superoperator formalism}
In order to treat discrete and continuous density operator
representations on an equal footing, we introduce the following
superoperator formalism. The set of linear operators acting on a
D-dimensional Hilbert space H is a $D^2$-dimensional complex
vector space ${\cal L(H)}$. Let us introduce operator "kets"
$\mid A) = A$ and "bras" $(A\mid = A^{\dagger}$, distinguished
from vector kets and bras by the use of round brackets. Then the
natural inner product on L(H), the trace-norm inner product, can
be written as $(A\mid B) = tr(A^{\dagger} B)$. The notation $S =
\mid A)(B\mid$ defines a superoperator S acting like \be S\mid X)
= \mid A)(B\mid X) = tr(B^{\dagger}X)A . \ee Now let the set
$\{\mid N_j)\}$ constitute a (complete or overcomplete) operator
basis; i.e., let the operator kets $\mid N_j )$ span the vector
space L(H). It follows that the superoperator ${\cal G}$ defined
by \be {\cal G}  \equiv\sum_{j}\mid N_j )(N_j \mid \ee is
invertible. The operators \be Q_{j}\equiv {\cal G}^{-1}\mid
N_{j}) \ee form a dual basis, which gives rise to the following
resolutions of the superoperator identity: \be\label{sup3} 1
=\sum_{j} \mid Q_{j} )(N_{j}\mid =\sum_{j} \mid N_{j}
)(Q_{j}\mid.  \ee An arbitrary operator A can be expanded as
\be\label{sup1} A =\sum_{j} \mid N_{j} )(Q_{j}\mid A) =\sum_{j}
N_{j}tr(Q_{j}^{\dagger} A) \ee and \be\label{sup2} A =\sum_{j}
\mid Q_{j} )(N_{j}\mid A) =\sum_{j} Q_{j}tr(N_{j}^{\dagger} A)
\ee These expansions are unique if and only if the operators
$N_{j}$ are linearly independent\cite{schack}.
\section{Projection method  as a semidefinite programming and finite quantum tomography}
\subsection{Bases and frames}\label{baseframe} In this section we
collect some rudimentary facts that will be used in what follows.

A {\it basis} is one of the most fundamental concepts in linear
algebra.

A set of linearly independent vectors $\{e_{i}\}_{i=1}^{n}$ in a
finite dimensional complex vector space $V$ is a {\it basis} for
$V$ if, for each $f\in V$, there exist coefficients  $c_{1},
c_{2}, ..., c_{n} \in {\cal C}$ such that \be
f=\sum_{i=1}^{n}c_{i} e_{i}.\ee The independence condition implies
that the coefficients $c_{1},...,c_{n}$ are unique.

For infinite dimensional vector spaces, the concept of a basis is
more complicated.

An $\{e_{i}\}_{i=1}^{\infty}\subseteq {\cal H}$ is an {\it
orthonormal system} (ONS) \cite{Christensen} if \be <e_{i},
e_{j}>=\delta_{ij}.\ee An ONS $\{e_{i}\}_{i=1}^{\infty}$ is an
{\it orthonormal basis} (ONB) if \be {\cal
H}=\bar{span}\{e_{i}\}_{i=1}^{\infty}\ee when
$\{e_{i}\}_{i=1}^{\infty}$ is an ONB, each $f\in {\cal H}$ can be
written as \be f=\sum_{i=1}^{\infty} <f, e_{i}> e_{i}. \ee

{\bf Definition}: Two sequences $\{x_{i}\}$ and $\{y_{i}\}$ in a
Hilbert space ${\cal H}$ are said to be {\it biorthonormal}, if
\be <x_{i}, y_{j}>=\delta_{ij}.\ee

A  sequence  $\{y_{i}\}$ biorthogonal to a basis $\{x_{i}\}$ for
${\cal H}$ is itself a basis for ${\cal H}$, and we have for each
$x$ the representation \be x=\sum_{i=1}^{\infty} <x, y_{i}>
x_{i},\; \mbox{and}\; x=\sum_{i=1}^{\infty} <x, x_{i}> y_{i}.  \ee
{\bf Frame}:  A family of elements  $\{f_{i}\}_{i\in I} \subseteq
{\cal H } $ is called a {\it frame} for ${\cal H}$ if there exist
constants $A, B> 0$ such that \be\label{frame12} A||f||^{2} \leq
\sum_{i\in I} |<f, f_{i}>|^{2} \leq B||f||^{2}, \forall f\in {\cal
H}, \ee where $I$ is a countable index set. The numbers $A, B$ are
called frame bounds. They are not unique. The optimal frame bounds
are the biggest possible value for A and the smallest possible
value for B in (\ref{frame12}). If we can choose A = B, the frame
is called tight. If a frame ceases to be a frame when any element
is removed, the frame is said to be exact. Since a frame
$\{f_{i}\}_{i\in I}$ is a Bessel sequence, the operator \be
T:l^{2}(I)\rightarrow {\cal H}\;\;,\;\; T\{c_{i\;i\in
I}\}=\sum_{i\in I}c_{i}f_{i}, \ee is bounded and linear; T is
sometimes called the preframe operator. The adjoint operator is
given by \be T^{\ast}:{\cal H}\rightarrow l^{2}(I) \;\;,\;\;
T^{\ast}f=\{<f,f_{i}>\}_{i=1}^{\infty}. \ee By composing the
operators T and $T^{\ast}$, we obtain the operator \be S:{\cal
H}\rightarrow {\cal H },\;\; Sf=TT^{\ast}f=\sum_{i=1}^{\infty}<f,
f_{i}> f_{i},\ee where  $S$ is called the frame operator with
\be\label{framebound} AI\leq S \leq BI.\ee The frame operator is
a bounded, positive, and invertible  operator.

 \subsection{Frames in finite-dimensional spaces}
We investigate the properties of a frame generated by a finite
subset of a Hilbert space.

 Calculation of the frame coefficients $\{<f, S^{-1}f_{i}>\}$
 involves inversion of the frame operator $S$. In practice it can be
 a problem if the underlying Hilbert space is infinite
 dimensional. There is an approach to the problem as follows
 \cite{Christensen}:

 Given the frame $\{f_{i}\}_{i=1}^{\infty}$ we consider finite
 subsets $\{f_{i}\}_{i=1}^{n}, n\in N$. It can be shown that
 $\{f_{i}\}_{i=1}^{n}$ is a frame for ${\cal
 H}_{n}=span\{f_{i}\}_{1}^{n}$ and the corresponding frame
 operator is $S_{n}:{\cal H}_{n}\rightarrow {\cal H}_{n}$ and the
 orthogonal projection $P_{n}$ on ${\cal H}_{n}$ is \be P_{n} f =\sum_{i=1}^{n} <f, S_{n}^{-1} f_{i}> f_{i}, \;\; f\in {\cal H}.\ee
 For $n\rightarrow \infty $, $P_{n}f\rightarrow f=\sum_{i=1}^{\infty} <f, S^{-1} f_{i}>
 f_{i}$, one can hope that the coefficients $<f, S_{n}^{-1}
 f_{i}>$converges to the frame coefficients for $f$, i.e., that
 \be\label{proj}<f, S_{n}^{-1}
 f_{i}>\rightarrow <f, S^{-1}
 f_{i}> \;\; \mbox{as}\; n\rightarrow \infty, \forall i\in I,\;\forall f\in {\cal H}.  \ee
 If (\ref{proj}) is satisfied we say that the projection method works. In this case the frame coefficients can be approximated
as close as we want using finite dimensional methods, i.e., linear
algebra, since $S_{n}$ is an operator on the finite dimensional
space ${\cal H}_{n}$. This is a very important property for
applications: for example, it makes it possible to use computers
to approximate the frame coefficients.

In the following subsection we obtain finite quantum tomography
using semidefinite
 programming.
\subsection{Finite quantum tomography via semidefinite programming} Quantum state
reconstruction schemes can be understood as an a posterior
estimation of density operator of a given quantum mechanical
system based on data obtained with the help of a macroscopic
measurement apparatus. Only if an infinite ensemble is given can
one find out the state. But infinite ensembles don't exist in
practice.  In a laboratory and in practice, we always deal with
finite ensembles of copies of the measured system. This implies
the need of developing novel tools specially designed to process
realistic and finite experimental samples. Then it is necessary
to truncate the Hilbert space to a finite dimensional basis
\cite{Vbuzek}.

Now in this work using the projection method and semidefinite
programming we express the mathematical structure correspond to
finite tomography and obtain the tomographic formula based on
finite Banach space.

At first  from (\ref{sup1})or (\ref{sup2}) we assume that \be
\rho=\sum_{j} \mid Q_{j} )(N_{j}\mid \rho) =\sum_{j}
Q_{j}tr(N_{j}^{\dagger} \rho) \ee is  a density matrix in infinite
dimensional  Banach space, where $\{N_{j}\}$ constitute a
operator basis in superoperator formalism. Also let
\be\label{supp1} \rho^{n}=\sum_{j=1}^{n} \lambda_{j}\mid N_{j}
)\ee be a density matrix in finite dimensional banach space which
is obtained from truncating the infinite dimensional Banach space.

Using  the properties  of density matrix we have \be\rho-\rho^{n}
\geq 0, \ee which  in comparison with semidefinite programming we
get $$F_{0}=\rho,\;\; F_{j}=\mid N_{j} )\;\; and \;\;
x_{j}=\lambda_{j},\;\; for \;\; j=1,...,n. $$

If we use the complementary slackness condition, and for a
feasible  $(\hat{Z}, {\lambda_{j\;}}_{max}), \;\; for \;\;
j=1,...,n$, we have \be \hat{Z}(\rho-\rho^{n})=0,\ee or \be
\hat{Z}(\rho-\sum_{j=1}^{n}\lambda_{j}\mid N_{j}))=0.\ee Using
resolution of the superoperator identity (\ref{sup3}) we obtain
\be \sum_{i}\hat{Z}\mid N_{i})(Q_{i}\mid [\rho
-\sum_{j}\lambda_{j}\mid N_{j})]=0, \;\; for \;\; j=1,...,n \ee
Therefore, we have \be \sum_{i}(\hat{Z}\mid
N_{i})[(Q_{i}\mid\rho)-\lambda_{i}]=0\;\;,\;\;i=1,...n.\ee  It is
obvious that $ (\hat{Z}\mid N_{i})=0$ for $i>n$ then we conclude
that $\mid N_{i})\in ker{\hat{Z}}$. Then we obtain \be
\lambda_{i}=(Q_{i}\mid\rho)=tr[\rho N^{\dagger}_{i}].\ee

Therefore we obtain the tomography formula in finite dimensional
Banach space as the follow: \be\label{fin1}
\rho^{n}=\sum_{i=1}^{n} \mid N_{i})(Q_{i}\mid\rho)=\sum_{i=1}^{n}
\mid N_{i}) tr[\rho N^{\dagger}_{i}]. \ee In the following, we
will consider density matrix with orthogonal states of the form:
$$\rho=\sum_{j}^{\infty} tr(\rho \ket{\psi_{j}}\bra{\psi_{j}})
\ket{\psi_{j}}\bra{\psi_{j}}$$ where is  a density matrix in
infinite dimensional  Hilbert space. In the superoperator
formalism we can write \be \mid N_{j})=\mid
Q_{j})=\ket{\psi_{j}}\bra{\psi_{j}} \ee
 Also   let
$$\rho^{n}=\sum_{j=1}^{n}
\lambda_{j}\mid N_{j} )=\sum_{j=1}^{n}
\lambda_{j}\ket{\psi_{j}}\bra{\psi_{j}}$$ be a density matrix in
finite dimensional Hilbert space which is obtained from truncating
the infinite dimensional Hilbert space and $\ket{\psi}$ is an
orthogonal state.

Using  (\ref{fin1}) we obtain the tomography formula in finite
dimensional Hilbert space as the follow: \be\label{fin2}
\rho^{n}=\sum_{j=1}^{n} tr(\rho \ket{\psi_{j}}\bra{\psi_{j}})
\ket{\psi_{j}}\bra{\psi_{j}}. \ee In the following we describe
some examples for finite dimensional quantum tomogarphy.
\section{Some examples for finite quantum tomography with semidefinite programming}
\subsection{Qudit tomography} We begin with the set of Hermitian generators of
SU(D); the generators, denoted by $\lambda_{j}$ , are labeled by a
Roman index taken from the middle of the alphabet, which takes on
values $j = 1, . . . ,D^2 - 1$ \cite{Rungta}. We represent the
generators in an orthonormal basis $|a>$, labeled by a Roman
letter taken from the beginning of the alphabet, which takes on
values $ a = 1, . . . ,D$. With these conventions the generators
are given by
\begin{eqnarray}
&\mbox{}&
j=1,\ldots,D-1:\nonumber\\
&\mbox{}&\hphantom{j=1,\ldots,}
   \lambda_j=\Gamma_a\equiv{1\over\sqrt{a(a-1)}}
   \left(\sum_{b=1}^{a-1}
   |b\rangle\langle b|-(a-1)|a\rangle\langle a|\right)
   \;,\quad 2\leq a\leq D\;,\label{eq:diagonal}\\
&\mbox{}&
j=D,\ldots,{(D+2)(D-1)/2}:\nonumber\\
&\mbox{}&\hphantom{j=1,\ldots,}
    \lambda_j=\Gamma_{ab}^{(+)}\equiv{1\over\sqrt2}
    \left(|a\rangle\langle b|+|b\rangle\langle a|\right)
    \;,\quad 1\leq a<b\leq D\;,\label{eq:plus}\\
&\mbox{}&
j=D(D+2)/2,\ldots,D^2-1:\nonumber\\
&\mbox{}&\hphantom{j=1,\ldots,}
    \lambda_j=\Gamma_{ab}^{(-)}\equiv{-i\over\sqrt2}
    \left(|a\rangle\langle b|-|b\rangle\langle a|\right)
    \;,\quad 1\leq a<b\leq D\;.\label{eq:minus}
\end{eqnarray}
In Eqs.(\ref{eq:plus}) and (\ref{eq:minus}), the Roman index $j$
stands for the pair of Roman indices, $ab$, whereas in
Eq.(\ref{eq:diagonal}), it stands for a single Roman index $a$.
The generators are traceless and satisfy
\begin{equation}
\lambda_j\lambda_k={1\over
D}\delta_{jk}+d_{jkl}\lambda_l+if_{jkl}\lambda_l\;.
\end{equation}
Here and wherever it is convenient throughout this paper, we use
the summation convention to indicate a sum on repeated indices.
The coefficients $f_{jkl}$, the structure constants of the Lie
group SU($D$), are given by the commutators of the generators and
are completely antisymmetric in the three indices.  The
coefficients $d_{jkl}$ are given by the anti-commutators of the
generators and are completely symmetric.

By supplementing the $D^2-1$ generators with the operator
\begin{equation}
\lambda_0\equiv{1\over\sqrt D}I\;,
\end{equation}
where $I$ is the unit operator, we obtain a Hermitian operator
basis for the space of linear operators in the qudit Hilbert
space.  This is an orthonormal basis, satisfying
\begin{equation}
{\rm tr} (\lambda_\alpha\lambda_\beta)=\delta_{\alpha\beta}\;.
\end{equation}
Here the Greek indices take on the values $0,\ldots,D^2-1$;
throughout this paper, Greek indices take on $D^2$ or more
values.  Using this orthonormality relation, we can invert
Eqs.(\ref{eq:diagonal})-(\ref{eq:minus}) to give
\begin{eqnarray}
|a\rangle\langle a|&=& {I\over D}+{1\over\sqrt{a(a-1)}}
\left(-(a-1)\Gamma_a+\sum_{b=a+1}^D\Gamma_b\right)\;,
\label{eq:jkdiagonal}\\
|a\rangle\langle b|&=&
{1\over\sqrt2}(\Gamma_{ab}^{(+)}+i\Gamma_{ab}^{(-)})
\;,\quad 1\leq a<b\leq D\;,\label{eq:jkplus}\\
|b\rangle\langle a|&=&
{1\over\sqrt2}(\Gamma_{ab}^{(+)}-i\Gamma_{ab}^{(-)}) \;,\quad
1\leq a<b\leq D\;.\label{eq:jkminus}
\end{eqnarray}

Any qudit density operator can be expanded uniquely as
\begin{equation}
\label{eq:rhoex} \rho= {1\over D}c_\alpha\lambda_\alpha\;,
\end{equation}
where the (real) expansion coefficients are given by
\begin{equation}
c_\alpha=D{\rm tr}(\rho\lambda_\alpha)\;.
\end{equation}
Normalization implies that $c_0=\sqrt D$, so the density operator
takes the form
\begin{equation}
\label{eq:rhoexp}
 \rho={1\over D}
\left(I+c_j\lambda_j\right) ={1\over D} (I+\vec c\cdot\vec\lambda)
\;.
\end{equation}
Here $\vec c=c_j\vec e_j$ can be regarded as a vector in a
$(D^2-1)$-dimensional real vector space, spanned by the
orthonormal basis $\vec e_j$, and $\vec\lambda=\lambda_j\vec e_j$
is an operator-valued vector.

In order to treat discrete density operator representation for a
qudit we introduce the superoperator formalism and SDP method.
Consider a discrete set of projection operators \cite{Rungta}
define in finite dimensional Banach space  \be
N_{\overrightarrow{n_{\alpha}}}=\mid
\overrightarrow{n_{\alpha}}><\overrightarrow{n_{\alpha}}\mid=\frac{1}{D}(1+\overrightarrow{\lambda}.\overrightarrow{n_{\alpha}})\;\;,\;\;\alpha=1,...,K.
\ee The corresponding superoperator,
\begin{equation}
{\cal G}=\frac{K}{D(D+1)}\!\left((D+1){|I)(I|\over D}+{\cal
T}\right)\;, \label{eq:Gdiagonal}
\end{equation}
where, orthonormal eigenoperators of ${\cal G}$ are
$\lambda_0=I/\sqrt D$ and ${\cal T}= \sum_{j}\mid
\lambda_{j}><\lambda_{j}\mid$.

We are now prepared to write the inverse of ${\cal G}$ with
respect to the left-right action as
\begin{equation}
{\cal G}^{-1}=\frac{D(D+1)}{K}\!\left({1\over D+1}{|I)(I|\over
D}+{\cal T}\right).
\end{equation}
Thus the dual operators are given by
\begin{equation}
\label{eq:gri} |Q_{n_{\alpha}})={\cal
G}^{-1}|N_{n_{\alpha}})=\frac{D(D+1)}{K}(\mid
N_{\alpha})-\frac{\mid I) }{D+1}).
\end{equation}
Using SDP method we get \be
F_{0}=\frac{1}{D}(1+c.\lambda)\;\;,\;\;f_{\alpha}=\mid
N_{\alpha})\;\; \mbox{and}\;\;\;
x_{\alpha}=\Lambda_{\alpha}\;\;\mbox{for}\;\; \alpha=1,...,K. \ee
From cmplementary slackness condition we have \be
\Lambda_{\alpha}=(Q_{n_{\alpha}}\mid\rho). \ee Therefore,
tomography relation in finite dimensional Banach space can be
represented in the form \be \rho^{K}=\sum_{\alpha=1}^{K}\mid
N_{n_{\alpha}})(Q_{n_{\alpha}}\mid\rho)=\sum_{\alpha=1}^{K}
Tr[Q_{n_{\alpha}}^{\dagger}\rho]N_{n_{\alpha}}=\frac{D(D+1)}{K}\sum_{\alpha=1}^{K}N_{\alpha}(I+Tr[N_{n_{\alpha}}\rho]).\ee
A qubit is two-level system, for which D = 2. There is a
one-to-one correspondence between the pure states of a qubit and
the points on the unit sphere, or Bloch sphere\cite{schack}. Any
pure state of a qubit can be written in terms of the Pauli
matrices $(\sigma_1,\sigma_2,\sigma_3)$, as
$$N_{\overrightarrow{n}}=\mid\overrightarrow{n}><\overrightarrow{n}\mid$$
where $\overrightarrow{n} = (n_1; n_2; n_3)$ is a unit vector, and
1 denotes the unit matrix. An arbitrary state $\rho$, mixed or
pure, of a qubit can be expressed as\be \rho = \frac{1}{2}(1 +
\overrightarrow{S}.\overrightarrow{\sigma})\ee where $0 \leq \mid
S\mid \leq 1$. \\In order to treat discrete density operator
representation for a qubit we introduce the superoperator
formalism and SDP method. Consider a discrete set of projection
operators\cite{schack} in superoperator formalism \be
N_{\overrightarrow{n_{\alpha}}}=\mid
\overrightarrow{n_{\alpha}}><\overrightarrow{n_{\alpha}}\mid=\frac{1}{2}(1+\overrightarrow{\sigma}.\overrightarrow{n_{\alpha}})\;\;,\;\;\alpha=1,...,K.
\ee The corresponding superoperator, \be\label{finit4} {\cal G}=
\sum_{\alpha=1}^{K}\mid N_{\overrightarrow{n_{\alpha}}})(
N_{\overrightarrow{n_{\alpha}}}\mid=\frac{1}{4}[K \mid
1)(1\mid+\sum_{\alpha}[\overrightarrow{n_{\alpha}}.\mid
\overrightarrow{\sigma})( 1\mid+\mid 1)(
\overrightarrow{\sigma}\mid
.\overrightarrow{n_{\alpha}}]+\sum_{j,k}\mid\sigma_{j})(\sigma_{k}\mid\sum_{\alpha}(n_{\alpha})_{j}(n_{\alpha})_{k}],\ee
generates dual-basis operators and expansion coefficients
proportional to those for the continuous representation
\cite{schack} if and only if\be
0=\sum_{\alpha}\overrightarrow{n}_{\alpha}\ee
$$\frac{1}{3}\delta_{jk}=\frac{1}{K}\sum_{\alpha}(n_{\alpha})_{j}(n_{\alpha})_{k}.$$
When these conditions are satisfied, the superoperator
(\ref{finit4}) simplifies to \be {\cal G}=\frac{K}{4}[\mid
1)(1\mid +\frac{1}{3}\sum_{j}\mid \sigma_{j})(\sigma_{j}\mid],\ee
with an inverse \be {\cal G}^{-1}=\frac{1}{K}[\mid 1)(1\mid
+3\sum_{j}\mid \sigma_{j})(\sigma_{j}\mid],\ee which generates
dual-basis operators \be Q_{\overrightarrow{n}_{\alpha}}={\cal
G}^{-1}\mid
N_{\overrightarrow{n}_{\alpha}})=\frac{1}{K}(1+3\overrightarrow{\sigma}.\overrightarrow{n}_{\alpha}).
\ee Then the density matrix in finite dimensional Banach space is
given by (\ref{supp1}). Using SDP method we get \be
F_{0}=\frac{1}{2}(1+S.\sigma)\;\;,\;\;f_{\alpha}=\mid
N_{\alpha})\;\; \mbox{and}\;\;\;
x_{\alpha}=\lambda_{\alpha}\;\;\mbox{for}\;\; \alpha=1,...,K. \ee
From cmplementary slackness condition we have \be
\lambda_{\alpha}=(Q_{n_{\alpha}}\mid\rho)=Tr[\frac{1}{2K}(1+3\overrightarrow{\sigma}.\overrightarrow{n}_{\alpha})(1+\overrightarrow{S}.\overrightarrow{\sigma})=\frac{1}{K}(1+3\overrightarrow{S}.\overrightarrow{n}_{\alpha})
)]. \ee Therefore, tomography relation (\ref{fin1}) in finite
dimensional Banach space can be represented in the form \be
\rho^{K}=\sum_{\alpha=1}^{K}\mid
Q_{\alpha})(N_{\alpha}\mid\rho)=\frac{1}{K}\sum_{\alpha=1}^{K}N_{\alpha}(1+3\overrightarrow{S}.\overrightarrow{n}_{\alpha}),\ee
For M qubits, we define the pure-product-state projector \be
N(\alpha) = N_{\alpha_{1}} \otimes ... \otimes N_{\alpha_{M}} =
\frac{1}{2^M}(1 + s · n_{\alpha_{1}})\otimes ... \otimes (1 + s .
n_{\alpha_{M}}) , \ee and \be Q(\alpha) =
Q_{n_{\alpha_{1}}}\otimes ... \otimes Q_{n_{\alpha_{M}}} =
\frac{1}{4\pi}^{M}(1 + 3s \cdot n_{\alpha_{1}}) \otimes ...
\otimes (1 + 3s \cdot n_{\alpha_{M}}),\ee where n stands for the
collection of unit vectors $n_{1}, . . . , n_{M}$. Any M-qubit
density operator can be expanded as \be
\rho^{K}=\sum_{\alpha=1}^{K}\mid
N_{\alpha})(Q_{\alpha}\mid\rho)=\frac{1}{K^M}\sum_{\alpha_{1},...,\alpha_{M}=1}^{K}N_{\alpha_{1}}(1+3\overrightarrow{S}.\overrightarrow{n}_{\alpha_{1}})\otimes...N_{\alpha_{M}}(1+3\overrightarrow{S}.\overrightarrow{n}_{\alpha_{M}}),\ee
where thus obtained result is in agreement with those of already
obtained by one of the authors in \cite{schack,Buzek}.

\subsection{Phase tomography} one possible means of describing the
phase of a quantum mechanical fields is in terms of the
Pegg-Barnett hermitian phase operator $\hat{\Phi}$
\cite{Pegg,Barnett,Buzek}. This operator is defined in a finite
(but arbitrary large) dimensional Hilbert space. In a
(s+1)-dimensional Hilbert space the phase state are defined as
\be\label{phase1} \mid
\theta>=\frac{1}{\sqrt{s+1}}\sum_{n=0}^{s}e^{in\Phi}\mid n>, \ee
this Hilbert space is spanned by a complete orthonormal set of
basis phase state $\mid \theta_{m}>$, given by (\ref{phase1})
with \be \theta_{m}=\theta_{0}+\frac{2\pi m}{s+1}\;\;,\;\;
m=0,1,...,s, \ee where $\theta_{0}$ is a reference phase. In
terms of the state $\mid \theta_{m}>$ the Hermitian phase
operator is \be\label{phase2}
\hat{\Phi}_{\theta}=\sum_{m=0}^{s}\theta_{m}\mid \theta_{m}><
\theta_{m}\mid. \ee From the definition of the phase state
(\ref{phase1}), we can express the projector $\theta_{m}\mid
\theta_{m}>< \theta_{m}\mid$ in terms of the number state basis:
\be \mid \theta_{m}>< \theta_{m}\mid= (s+1)^{-1}
\sum_{n=0}^{s}\sum_{n^{\prime}=0}^{s}e^{i(n^{\prime}-n)\Phi}\mid
n^{\prime}><n\mid.\ee

In this case $\Phi_{\theta}$ is orthonormal then we can write the
tomography using semidefinite programming.

At first we assume that \be\rho=\int_{\theta} Tr(\rho
\Phi_{\theta})\Phi_{\theta}d\mu_{\theta}, \ee is a density matrix
in infinite dimensional Banach space. Also let \be
\rho^{\prime}=\sum_{\theta}\lambda_{\theta} \mid
\hat{\Phi_{\theta}}), \ee be a density matrix in finite
dimensional Banach space which is obtained from truncating the
infinite dimensional Banach space. Using the properties of
density matrix we have \be \rho-\rho^{\prime}\geq 0, \ee which is
comparison with semidefinite programming we get \be
F_{0}=\rho\;\;,\;\; F_{\theta}=\mid \hat{\Phi}_{\theta})\;\;\;
\mbox{and}\; x_{\theta }=\lambda_{\theta}\;\;,\;\;\mbox{for}\;\;
\theta=\theta_{0},...,\theta_{0}+2\pi. \ee If we use the
complementary slackness  condition, and for a feasible
($\hat{Z},{\lambda_{\theta}}_{max}$), for
$\theta=\theta_{0},...,\theta_{0}+2\pi$, we have \be
\hat{Z}(\rho-\rho^{\prime})=0\;\;\mbox{or}\;\;\hat{Z}(\rho-\lambda_{\theta}
\mid \hat{\Phi}_{\theta}))=0. \ee Similar to superoperator
formalism we obtain \be \lambda_{\theta}=(\hat{\Phi}_{\theta}\mid
\rho )=Tr[\rho\hat{\Phi}_{\theta}].  \ee Therefore we obtain the
tomography formula in finite dimensional Hilbert space as the
follow:\be \rho^{\prime}=\sum_{\theta}\mid
\hat{\Phi_{\theta}})(\hat{\Phi_{\theta}}\mid\rho )=
\sum_{\theta}\mid \hat{\Phi_{\theta}})Tr[\rho \hat{\Phi}_{\theta}
].\ee If we generalized it when $\theta $ is continuous, in this
case we have
 \be \rho=\int_{\theta_{0}}^{\theta_{0}+2\pi}Tr[\rho \hat{\Phi_{\theta}}]\hat{\Phi}_{\theta} d\theta. \ee
Using (\ref{phase2}) $Tr[\rho \hat{\Phi_{\theta}}]$ obtain as
follows \be Tr[\rho \hat{\Phi}_{\theta}]=
Tr[\rho\sum_{m=0}^{s}\theta_{m}\mid \theta_{m}>< \theta_{m}\mid
]= 2\pi \sum_{m}\theta_{m}\frac{1}{s+1}P_{PB}(\theta)_{m}, \ee
where $P_{PB}$ is probability of measuring a particular value of
phase and is normalized so that the integral of
$P_{PB}(\Phi_{\theta})$ over a $2\pi$ region of $\theta$ is equal
to one. \be
P_{PB}(\Phi_{\theta})=\frac{1}{2\pi}\sum_{n,n^{\prime}=0}^{s}e^{i(n^{\prime}-n)\phi}<n\mid
\rho \mid n^{\prime}>=<\theta\mid\rho\mid \theta>, \ee where thus
obtained results are in agreement with those of already obtained
by one of the authors in \cite{Pegg,Barnett,Buzek}.

 A very
important subset of these states will be the physical partial
phase states, of which the coherent state is a particular
example. The phase states are themselves unphysical and so the
best attempt at a physical phase measurement will only project
the system into a physical partial phase state \cite{Pegg}. In
the following, we obtain a physical partial phase state tomography
i.e., coherent spin states tomography.
\subsection{Coherent spin states tomography}
To reconstruct a mixed or pure quantum state of a spin  s  is
possible through coherent states: its density matrix  is fixed by
the probabilities to measure the value  s along 4s(s+1)
appropriately chosen directions in space. Thus, after inverting
the experimental data, the statistical operator is parameterized
entirely by expectation values.

A coherent spin state $\mid n >$ is associated to each point of
the surface of the unit sphere. \be \mid n >\equiv exp[-i\theta
m(\phi).\hat{s}]\mid s,n_{z}>, \ee where
$m(\phi)=(-sin\phi,cos\phi,0)$.

A stereographic projection of the surface of the sphere to the
complex plane give the expansion of a coherent state
\cite{Weigert} as follows \be \mid s,n>=\frac{1}{(1+\mid z\mid^{2}
)^{s}}\sum_{k=0}^{2s}\left(\begin{array}{c}2s//k
\end{array}\right)^{1/2} z^{k}\mid s-k,n_{z}>. \ee
In order to show that the density matrix $\rho$ of a spin s is
determined unambiguously by appropriate measurement with a
Stern-Gerlach apparatus one precedes as follows. Distribute
$N_{s}=(2s+1)^{2}$ axes $\mid s,n>$ with $1\leq n\leq N_{s}$,
over (2s+1) cones about the z axis with different opening angles
such that the set of the (2s+1) directions on each cone is
invariant under a rotation about  z  by an angle
$\frac{2\pi}{(2s+1)}$.

An unnormalized statistical density operator  is then fixed by
measuring the $N_s$ relative frequencies \be p_{n}(n_{n}) =<
n_{n}\mid\rho\mid n_{n}>\;\;,\;\; 1 \leq n \leq N_{s}, \ee that
is, by the expectation values of the statistical operator
$\hat{\rho}$ in the coherent states $\mid n_{n}>$. You obtain
$N_s$ linear relations between probabilities $P_{n}(n_{n})$ and
the matrix elements of the density matrix with respect to the
basis $|s - k, n_z>$. This set of equations can be inverted by
standard techniques if the directions $n_{n}$ are chosen as
described above. For a spin s, the projection operators
\be\label{project1} \mid Q_{n})=\mid n_{n}><n_{n} \mid, \ee
constitute thus a quorum Q. In general, a quorum is defined as a
collection of (hermitian) operators having the property that their
expectation values are sufficient to reconstruct the quantum
state of the system at hand. $(Q^{n}\mid $ defined as the dual of
the quorum (\ref{project1}): \be
\frac{1}{(2s+1)}\sum_{n=1}^{N_{s}}\sum_{n^{\prime}=1}^{N_{s}}\mid
Q_{n})(Q^{n^{\prime}}\mid= \delta_{n}^{n^{\prime}}\;\;,\;\; 1\leq
n, n^{\prime} \leq N_{s}. \ee
Therefore,  this coherent spin
state introduced above is same as the phase state.

In order to obtain spin tomography relation in the finite
dimensional Banach space we assume that \be\rho=\int Tr(\rho \mid
Q_{n}))(Q^{n}\mid d\mu_{n}, \ee is a density matrix in infinite
dimensional Banach space. Also let \be
\rho^{\prime}=\sum_{n}\lambda_{n} \mid \hat{Q}^{n}), \ee be a
density matrix in finite dimensional Banach space which is
obtained from truncating the infinite dimensional Banach space.
Using the properties of density matrix we have \be
\rho-\rho^{\prime}\geq 0, \ee which is comparison with
semidefinite programming  and using complementary slackness
condition, we get \be
\hat{Z}(\rho-\rho^{\prime})=0\;\;\mbox{or}\;\;\hat{Z}(\rho-\lambda_{n}
\mid \hat{Q}^{n}))=0. \ee Similar to supperoperator formalism we
obtain \be \lambda_{n}=(\hat{Q}_{n}\mid \rho
)=Tr[\rho\hat{Q}_{n}]= P_{n}.  \ee Therefore we obtain the
tomography formula in finite dimensional Hilbert space as the
follow:\be \rho^{s}=\frac{1}{2s+1}\sum_{n=1}^{N_{s}}P_{n}Q^{n},\ee
where the coefficients $P_{n}$ satisfy \be 0\leq P_{n} \leq
1\;\;,\;\; 1 \leq n \leq N_{s}. \ee The operators $Q_{n}$ do even
define an optimal quorum since exactly $(2s + 1)^{2}$ numbers
have to be determined experimentally which equals the number of
free real parameters of the (unnormalized) hermitian density
matrix $\hat{\rho}$ . Thus obtained results are in agreement with
those of already obtained by one of the authors in
\cite{Pegg,Barnett,Buzek,Weigert}.

It is important to note that, although each of the $P_n$ is a
probability, they do not sum up to unity: \be\label{project2} 0 <
\sum_{n=1}^{N_s} P_n < (2s + 1)^2\ee .  This is due to the fact
that they all refer to different orientations of the
Stern-Gerlach apparatus, being thus associated with the
measurement of incompatible observables,

\be\label{project3} [Q_n, Q_{n^{\prime}}]\neq 0 , 1 \leq n,
n^{\prime} \leq N_s ,\ee since the scalar product
$<n_n|n_n^{\prime}>$ of two coherent states is different from
zero. The sum in (\ref{project2}) cannot take the value $(2s +
1)^2$ since this would require a common eigenstate of all the
operators $Q_n$ which does not exist due to (\ref{project3}). By
an appropriate choice of the directions $n_n$ (all in the
neighborhood of one single direction $n_0$, say), the sum can be
arbitrarily close to $(2s+1)^2$ for states 'peaked' about $n_0$.
Similarly, the sum of all $P_n$ cannot take on the value zero
since this would require a vanishing density matrix which is
impossible. If, however, considered as a sum of expectation
values, there is no need for the numbers $P_n$ to sum up to unity.
Nevertheless, they are not completely independent when arising
from a statistical operator: its normalization implies that \be
Tr [ \rho^{s} ] =
Tr[\frac{1}{2s+1}\sum_{n=1}^{N_{s}}P_{n}Q^{n}]=1, \ee turning one
of the probabilities into a function of the $(2s + 1)^2 - 1 = 4s(s
+ 1)$ others, leaving us with the correct number of free real
parameters needed to specify a density matrix\cite{Weigert}.

\section{Conclusion}Using the
elegant method of convex semidefinite  optimization method and
superoperator formalism, we have been able to obtain the quantum
tomography in finite dimensional representation for some set of
mixed density matrices. In this method we have been able to
obtain finite qudit, N-qubit quantum tomography, phase tomography
and coherent spin state tomography , where these results that
obtained are in agreement with those of ref\cite{schack} and
\cite{Pegg,Barnett,Buzek,Weigert} .

\end{document}